\documentclass{article}
\usepackage{spconf,amsmath,graphicx, multirow}
\usepackage{booktabs}
\usepackage[usenames,dvipsnames]{color}
\usepackage{subcaption,amsfonts, tabularx, hyperref}

\newlength{\bibitemsep}
\newlength{\bibparskip}
\let\oldthebibliography\thebibliography
\renewcommand\thebibliography[1]{%
  \oldthebibliography{#1}%
  \setlength{\parskip}{\bibitemsep}%
  \setlength{\itemsep}{\bibparskip}%
}

\definecolor{turquoise}{cmyk}{0.65,0,0.1,0.3}
\definecolor{purple}{rgb}{0.65,0,0.65}
\definecolor{dark_green}{rgb}{0, 0.5, 0}
\definecolor{orange}{rgb}{0.8, 0.6, 0.2}
\definecolor{red}{rgb}{0.8, 0.2, 0.2}
\definecolor{darkred}{rgb}{0.6, 0.1, 0.05}
\definecolor{blueish}{rgb}{0.0, 0.3, .6}
\definecolor{light_gray}{rgb}{0.7, 0.7, .7}
\definecolor{pink}{rgb}{1, 0, 1}
\definecolor{greyblue}{rgb}{0.25, 0.25, 1}

\title{An empirical study on speech restoration guided by self-supervised speech representation}
\name{Jaeuk Byun, Youna Ji, Soo-Whan Chung, Soyeon Choe, Min-Seok Choi}
\address{NAVER Cloud Corporation, Seongnam, South Korea}
\begin{document}
%
\maketitle
\begin{abstract}

Enhancing speech quality is an indispensable yet difficult task as it is often complicated by a range of degradation factors. In addition to additive noise, reverberation, clipping, and speech attenuation can all adversely affect speech quality. Speech restoration aims to recover speech components from these distortions. This paper focuses on exploring the impact of self-supervised speech representation learning on the speech restoration task. Specifically, we employ speech representation in various speech restoration networks and evaluate their performance under complicated distortion scenarios. Our experiments demonstrate that the contextual information provided by the self-supervised speech representation can enhance speech restoration performance in various distortion scenarios, while also increasing robustness against the duration of speech attenuation and mismatched test conditions.

\end{abstract}
\section{Introduction}
\label{sec:intro}
Speech is an intuitive and efficient means of human communication. However, the speech signal is often affected by various environmental factors, such as noise, interference, room reverberation, low-quality devices, interruptions in acoustic channels, packet loss, and jitters. To mitigate these degradation factors, numerous studies have attempted to address single-distortion issues, such as denoising~\cite{demucs, se-conformer}, dereverberation~\cite{nn-wpe}, declipping~\cite{declipping}, and bandwidth extension~\cite{be-wavenet,nu-wave2}. Nevertheless, scenarios with multiple simultaneous distortions have received less attention, despite being the most practical case.

Speech restoration is a challenging task that aims to recover speech components from various degradation factors. Recent approaches employing deep learning techniques, such as adversarial training~\cite{voicefixer,hifi++,metricgan} and diffusion methods~\cite{universal, sgmse+}, have gradually addressed this problem by considering multiple distortion scenarios. Another approach to speech restoration is to leverage additional information from different modalities. Many methods have utilized linguistic representations from text and visual cues to predict missing speech components~\cite{speechpainter,morrone2021audio}. However, these methods typically require text transcriptions or video recordings corresponding to speech signals, which are often inaccessible.
In recent years, self-supervised learning (SSL) methods have shown promising results on various recognition tasks by leveraging speech representations from a massive amount of unlabelled data~\cite{wav2vec2, wavlm, data2vec}. Several studies have adopted this representation for speech enhancement tasks, indicating the potential to improve enhancement performance, primarily targeting the denoising task~\cite{bsse-se, watanabe-se, demucs-ssl}.

This paper examines the impact of SSL representation on speech restoration. We assume that SSL models, pre-trained for masked prediction, provide well-contextualized features that can be used as conditioning signals to restore speech. We evaluate the performance of these features on speech restoration tasks under various lengths of speech attenuation, out-of-domain generalization, and signal-to-noise (SNR) conditions. Our results demonstrate that incorporating SSL features dramatically improves the performance of existing generation networks, especially for long attenuation and unseen test conditions.

\section{Speech restoration with SSL representations}
\label{sec:ssl_model}

\begin{figure*}[bt]
    \centering
    \begin{subfigure}[b]{0.245\textwidth}
      \centerline{\includegraphics[width=1\textwidth]{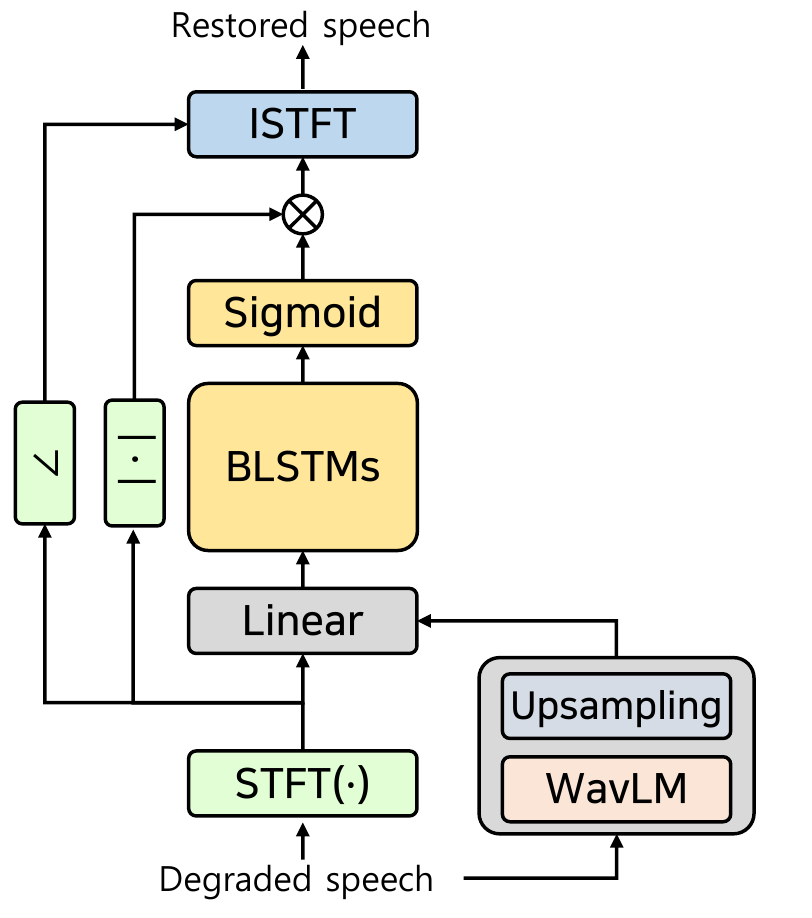}}
      \caption{BLSTM}
      \label{fig1a}
    \end{subfigure}
    \hspace{0.02\textwidth}
    \begin{subfigure}[b]{0.245\textwidth}
      \centerline{\includegraphics[width=1\textwidth]{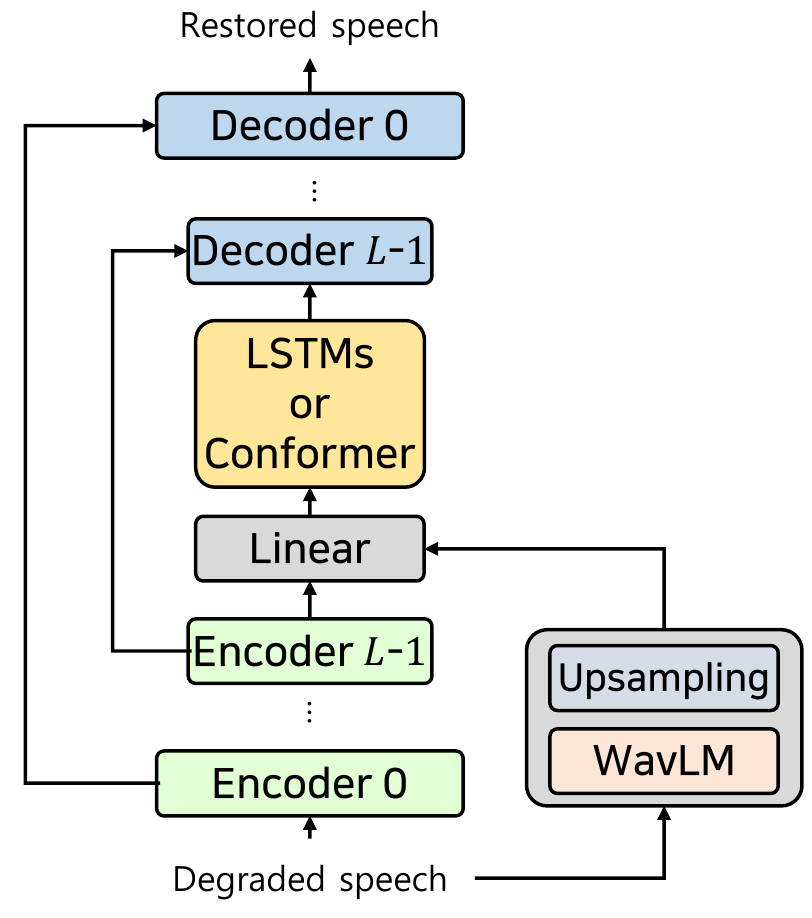}}
      \caption{DEMUCS and SE-Conformer}
      \label{fig1b}
    \end{subfigure}
    \hspace{0.02\textwidth}
    \begin{subfigure}[b]{0.245\textwidth}
      \centerline{\includegraphics[width=1\textwidth]{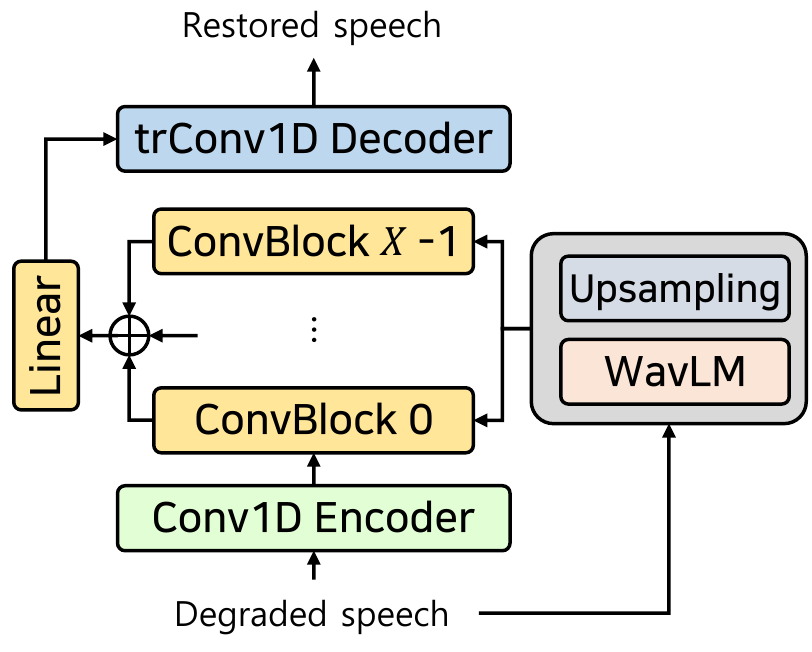}}
      \caption{Conv-TasNet}
      \label{fig1c}
    \end{subfigure}
    \caption{Structures for backbone network utilizing SSL representation.}
    \label{fig1}
    \vspace{-8pt}
\end{figure*}

\subsection{SSL based speech representation models}

Self-supervised learning methods aim to learn useful data representations without relying on human annotations. Recently, these methods have demonstrated remarkable results on various speech-related tasks such as automatic speech recognition~\cite{data2vec}, speaker recognition~\cite{wavlm}, and emotion recognition~\cite{wav2vec2}. One prominent example is the work by \cite{wav2vec2}, where the model is trained to predict the pseudo-labels of masked regions using unmasked speech parts. By leveraging this pre-training task, the model learns to capture the global structure and long-term dependencies of speech signals. Another noteworthy approach is WavLM \cite{wavlm}, a variant of HuBERT \cite{hubert} that jointly learns masked speech prediction and denoising during pre-training. This method has achieved remarkable performance on the SUPERB challenge~\cite{superb}.

\subsection{Problem formulation}
We expect that incorporating contextualized information into speech restoration, particularly in recovering speech components lost due to various degradation factors such as clipping, limited bandwidth, and attenuation, will be beneficial since training objectives in \cite{wavlm, data2vec, wav2vec} encourage models to capture the global structure of the signal \cite{bsse-se, watanabe-se}. We define the degraded speech signal $x$ as $x = f\left(s\right) + n \in {\mathbb{R}^{T}}$, where $f$ represents various degradation factors, and $s$ and $n$ are the original speech and noise signals, respectively. The goal of speech restoration is to find a mapping function $g:{\mathbb{R}^T} \to {\mathbb{R}^T}$ that transforms the degraded speech signal $x$ into a high-quality restored speech signal $\hat s$. To achieve this, we utilize the SSL representation as a conditioning signal for the speech restoration network $g$, and we formulate the problem as $\hat s = g\left( {x | h\left( x \right)} \right)$, where $h$ represents the SSL network.

\section{Network structures utilizing SSL representation}
\label{sec:model_description}

In this section, we present various approaches that utilize SSL representations for speech restoration. We focus on the most prominent models and provide detailed explanations in the following subsections. We evaluate their effectiveness in the experiments and conduct a comprehensive analysis of their performance in the next section.

\subsection{Spectral mask estimation}
In speech enhancement, spectral mask estimation in the time-frequency domain is a common approach that attenuates the magnitude spectrogram using the estimated ratio mask but neglects the distorted phase information. This method is effective for noise and interference that are modeled as additive relationships. In \cite{bsse-se}, SSL representations are incorporated into this mask estimation method as shown in Figure ~\ref{fig1a}. A simple bi-directional recurrent neural network (BLSTM) is designed where the SSL representation is injected before the BLSTM layer to integrate the degraded speech feature and its corresponding SSL feature. The model's noise suppression performance has improved compared to previous models.

\subsection{Raw waveform estimation}

In recent years, deep learning has enabled the restoration of signals in the waveform domain. These methods typically involve a 1-dimensional convolution layer with a large kernel and stride to simulate spectral analysis in STFT, while a transposed convolution layer is used to synthesize acoustic embedding into the waveform. This approach reduces the artifacts caused by STFT, resulting in signals with improved perceptual quality.

In \cite{demucs-ssl}, the authors used SSL representations on the DEMUCS structure to improve the overall quality of speech enhancement, demonstrating superiority over phonetic representations. We applied the conditioning method from \cite{demucs-ssl} to both DEMUCS and SE-Conformer models as shown in Figure ~\ref{fig1b}. Both models have the same U-Net-like structure except for the Conformer bottleneck in SE-Conformer. SSL representations are incorporated with distorted speech features before the bottleneck for both models, as this approach showed the best performance among methodologies.
In addition, we investigated the Conv-TasNet model for speech enhancement in the waveform, with several modifications for speech restoration based on \cite{convtasnet_synthesis}. In our approach, the SSL embedding is integrated with the inputs of the first convolution layer in every ConvBlock of Conv-TasNet, as illustrated in Figure ~\ref{fig1c}, where a ConvBlock is comprised of a point-wise convolution, a depth-wise convolution, and a residual connection.

\section{Experiments}
\label{sec:experiments}

\subsection{Dataset}
\label{subsec:dataset}

We used the VCTK-DEMAND dataset~\cite{vctk_demand}, which is a widely used benchmark for speech enhancement. The dataset comprises premixed noisy speech and includes a training set of 11,572 utterances from 28 speakers mixed with noise samples at 4 SNRs (0, 5, 10, and 15 dB). The test set contains 824 utterances from 1 male and 1 female speaker with 4 SNRs (2.5, 7.5, 12.5, and 17.5 dB). Speakers p286 and p287 were separated from the training set and used for validation as in \cite{demucs}. We degraded the VCTK-DEMAND test set by simultaneously applying several speech degradation factors, which are described in the next subsection. We evaluated speech quality using PESQ\footnote{\label{note1}https://github.com/schmiph2/pysepm}~\cite{pesq} and NISQA\footnote{https://github.com/gabrielmittag/NISQA}~\cite{nisqa}, and intelligibility using STOI\footref{note1}~\cite{stoi} with open-source toolkits.

\subsection{Speech degradation factors}
We evaluated 4 types of speech degradation factors: background noise, audio clipping, limited bandwidth, and speech attenuation. All these factors were applied simultaneously in all evaluations, unless otherwise stated.

\noindent\textbf{Background Noise}
We followed the specifications of the VCTK-DEMAND dataset described in Sec.~\ref{subsec:dataset}.

\noindent\textbf{Clipping}
It occurs when the input signal has excessive energy causing the audio codec to be unable to represent the amplitude adequately. To simulate clipping, we randomly limited the dynamic range of the amplitude from 0.06 to 0.9 in ratio.

\noindent\textbf{Limited bandwidth}
It is often caused by coding protocols, transmission channels, or room acoustics, leading to significant spectral corruption and a restricted frequency band range. To simulate limited bandwidth, we applied a low-pass filter with a random cutoff frequency between 2 to 8 kHz.

\noindent\textbf{Attenuation}
It can occur when the power of the speech signal is suddenly and significantly reduced, potentially due to interruptions in the acoustic channels. In our experiments, we attenuated speech amplitude by applying random gains ranging from 0 to 0.01 and random durations ranging from 10 to 50 ms to several (maximum 20) random regions.

\subsection{Implementation detail}

We trained all the compared models within our speech restoration framework, ensuring that our models obtained similar scores to those reported in \cite{demucs, bsse-se, demucs-ssl} using the original configurations. In our experiments, we adjusted the depth ($L$) of the U-Net structure to 4 to align the temporal resolution with SSL representations and set the number of channels in the first convolution layer to 48. For the remaining model parameters, we followed the original configurations.

To utilize SSL representation, we used the pre-trained WavLM-Large model~\cite{wavlm}, which has shown good results among compared SSL methods in~\cite{watanabe-se}. We followed the setup of SUPERB~\cite{superb} and froze the parameters of the SSL model. We took the weighted average of representations from all transformer encoder layers with learnable weights to yield the final SSL embedding. To match the frame rates of SSL with those of the intermediate representation of the backbone network, we repeated frames. For conditioning with SSL, we simply concatenated the intermediate features with the SSL representation, as we found no significant difference compared to applying \cite{film}.

\subsection{Results}
\label{sec:results}

\begin{table}[t]
\begin{tabular}{cccccc}\toprule

\multirow{2}{*}{Model}        & \multicolumn{2}{c}{Settings}                         & \multirow{2}{*}{PESQ} & \multirow{2}{*}{NISQA} & \multirow{2}{*}{STOI} \\\cmidrule{2-3}
                              & \multicolumn{1}{c}{Aug}  & SSL                       &                       &                        &                       \\ \hline
\multirow{3}{*}{BLSTM}        &                           &                           & 1.64                  & 2.18                   & 0.869                 \\ \cline{4-6} 
                              & \checkmark &                           & 1.86                  & 2.42                   & 0.882                 \\ \cline{4-6} 
                              & \checkmark & \checkmark & 2.00                  & 2.50                   & 0.892                 \\ \hline
\multirow{3}{*}{DEMUCS}       &                           &                           & 1.65                  & 2.11                   & 0.870                 \\ \cline{4-6} 
                              & \checkmark &                           & 2.52                  & 3.66                   & 0.926                 \\ \cline{4-6} 
                              & \checkmark & \checkmark & 2.59                  & 3.51                   & 0.930                 \\ \hline
\multirow{3}{*}{SE-Conformer} &                           &                           & 1.64                  & 2.22                   & 0.868                 \\ \cline{4-6} 
                              & \checkmark &                           & 2.61                  & 3.66                   & 0.929                 \\ \cline{4-6} 
                              & \checkmark & \checkmark & \bf{2.90}                  & 3.86                   & \bf{0.943}                 \\ \hline
\multirow{3}{*}{Conv-TasNet}   &                           &                           & 1.74                  & 2.26                   & 0.876                 \\ \cline{4-6} 
                              & \checkmark &                           & 2.60                  & 3.60                   & 0.930                 \\ \cline{4-6} 
                              & \checkmark & \checkmark & 2.79                  & \bf{3.87}                   & 0.936                 \\ \hline
\end{tabular}
\caption{Speech restoration results on VCTK-DEMAND with several degradation factors (noise, clipping, LPF, and attenuation). Various network structures are compared according to the additional augmentation and the use of SSL representation.}
\label{t1:overall}
\end{table}

\begin{table}[]
\fontsize{9pt}{12pt}
\selectfont
\begin{tabular}{ccccccc}\toprule
         & Measure & Noise & Clip  & LPF   & Att.  & All   \\ \midrule
\multirow{2}{*}{Base}   & PESQ    & 2.88  & 3.61  & 3.81  & 3.76  & 2.61  \\ 
                            & STOI    & 0.944 & 0.971 & 0.982 & 0.981 & 0.929 \\ 
\multirow{2}{*}{+SSL rep.}     & PESQ    & 3.08  & 3.77  & 3.96  & 3.93  & 2.90  \\ 
                            & STOI    & 0.952 & 0.978 & 0.988 & 0.987 & 0.943 \\ 
\bottomrule
\end{tabular}
\caption{Speech restoration results of SE-Conformer for single-distortion cases with and without SSL representation.}
\label{t2:bydistortion}
\end{table}

\subsubsection{Effect of model structure}

Table \ref{t1:overall} summarizes the speech restoration performance on VCTK-DEMAND with several degradation factors. For each model, we compared three progressive settings: 1) baseline configuration, 2) additional augmentation of training data with the speech degradation factors (clipping, limited bandwidth, and attenuation) described in Section 4.2, with probabilities of 0.25, 0.5, and 0.8, respectively, and 3) adding the condition to 2) with SSL representation. The results show that the baseline configuration is ineffective in dealing with unseen types of degradation factors, while it performs well when the degradation factors are given as additional augmentations during training, except for the masking-based BLSTM. This is because the masking approach only suppresses components in the input speech signal, mainly for denoising tasks, and is limited in recovering missing components. The other three waveform-generation models show moderate results, which are further improved when applied with SSL representation in most cases. Among the various model structures, SE-Conformer showed the most significant improvements, with PESQ improvement of \textbf{0.29}, NISQA improvement of \textbf{0.20}, and STOI improvement of \textbf{0.014} when using SSL representation. A possible explanation is that the global context captured by SSL representation may help to provide a better attention map in the Conformer blocks than the intermediate representation of the convolutional encoder alone, resulting in better signal estimation.

We also evaluated SE-Conformer with and without SSL representation under single distortion scenarios, each containing only one individual degradation factor, as shown in Table.~\ref{t2:bydistortion}. The results show that the use of SSL representation improves both PESQ and STOI for all individual degradation factors. Furthermore, the improvement is most pronounced in the scenario with multiple simultaneous distortions.

\subsubsection{Effect of duration for speech attenuation}

\begin{figure}[t]
    \centering
    \centerline{\includegraphics[width=0.86\columnwidth]{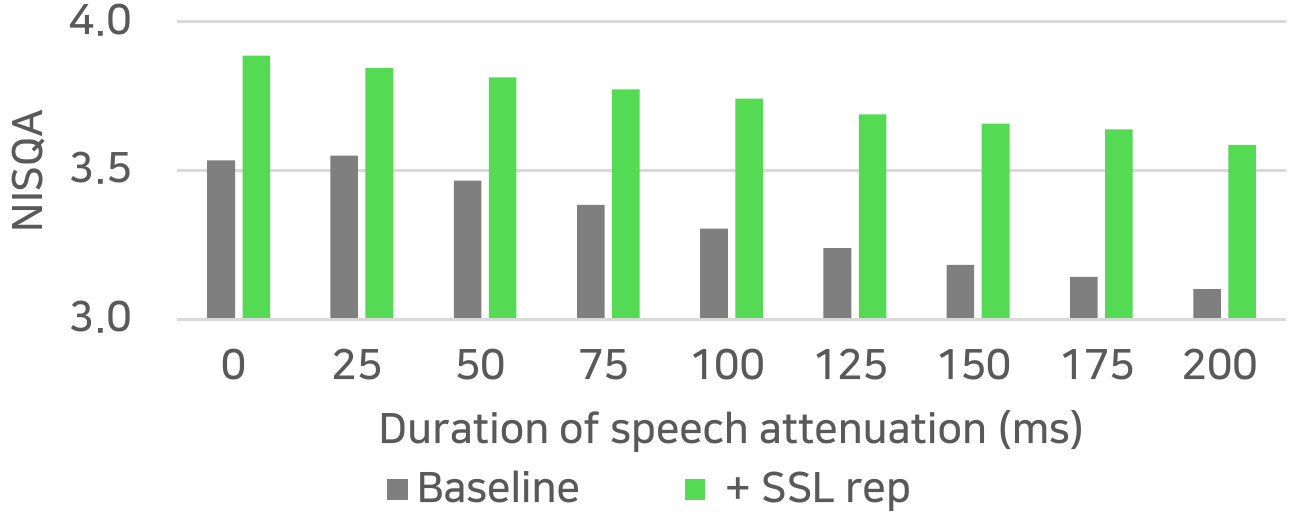}}
    \caption{Performance comparison in terms of NISQA using SE-Conformer with and without SSL representation according to the length of speech attenuation.}
    \label{fig2}
    \vspace{-8pt}
\end{figure}
We hypothesized that the SSL representation trained for masked prediction contains contextualized speech information, which could enhance the restoration network's robustness against longer duration of speech attenuation. To verify this hypothesis, we evaluated the SE-Conformer models with various lengths of attenuation from 0 to 200 ms, while keeping all other degradation parameters fixed. The NISQA scores are compared in Figure \ref{fig2}. The results suggest that the performance improvement by SSL representation becomes greater for longer speech attenuation. It is noteworthy that the models were trained using a database with only 10-50 ms of attenuation, and the test condition is harder and does not match the training data.

\subsubsection{Performance on mismatched data}

\begin{figure}[bt]
    \captionsetup[subfigure]{aboveskip=5pt}
    \centering
    \begin{subfigure}[b]{0.43\columnwidth}
      \centerline{\includegraphics[width=1\textwidth]{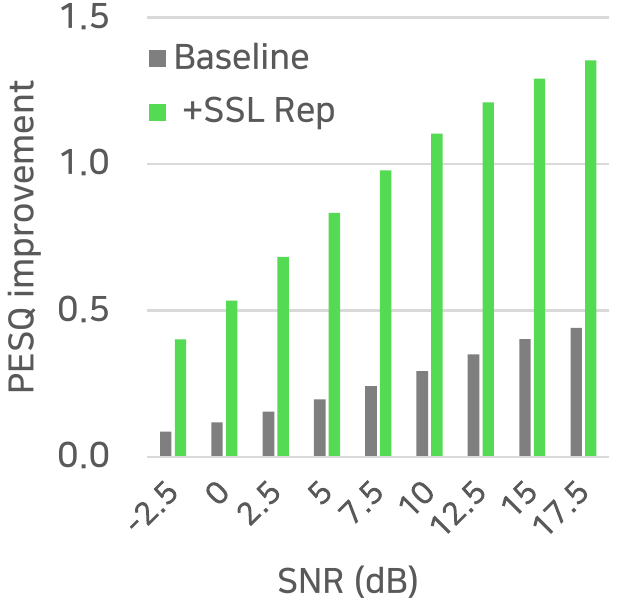}}
      \caption{PESQ}
      \label{fig3a}
    \end{subfigure}
    \hspace{0.03\columnwidth}
    \begin{subfigure}[b]{0.45\columnwidth}
      \centerline{\includegraphics[width=1\textwidth]{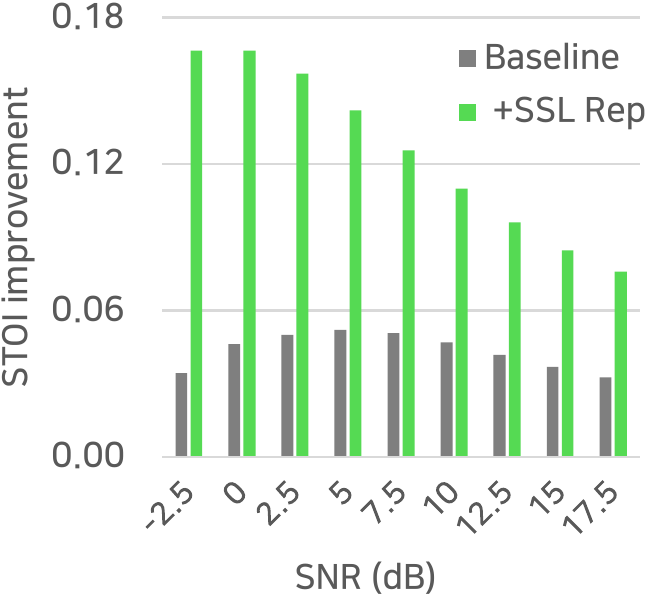}}
      \caption{STOI}
      \label{fig3b}
    \end{subfigure}
    \caption{Speech restoration performance on TIMIT-MUSAN (out of domain) dataset at various unseen SNRs using SE-Conformer models trained on VCTK-DEMAND dataset.}
    \label{fig3}
    \vspace{-8pt}
\end{figure}

SSL's generalization ability is a strong point, as it can be trained on a large amount of data without labels \cite{watanabe-se}. We evaluated the generalization ability of SSL representations for speech restoration by testing VCTK-trained SE-Conformer models on 1000 degraded mixtures from unseen data domains, using TIMIT \cite{timit} and MUSAN \cite{musan} datasets. The signals were mixed at 9 equally spaced SNR conditions in $\left[-2.5, 17.5\right]$. The evaluation focused on the PESQ and STOI improvements from the noisy mixture, which are summarized in Figure \ref{fig3}. 

In our experiments, we observed that incorporating SSL representation leads to more robust performance in mismatched conditions, as indicated by larger improvements in both PESQ and STOI compared to the baseline. The STOI improvements were particularly pronounced for lower SNRs. We found that many of the severely distorted samples in the baseline result were recovered by incorporating SSL representation, possibly due to better preservation of the phonetic structure with contextual information from SSL. However, the trend for PESQ was the opposite, with larger improvements in high SNR. This is likely because SSL representation lacks local and fine structure information that is necessary for high-quality restoration, as pointed out in \cite{bsse-se, watanabe-se}.

\section{Conclusions}
\label{sec:prior}

We investigated the effectiveness of SSL representation on speech restoration problem with multiple simulated distortions. We compared one spectral mask estimation and three waveform generation models conditioned on SSL representation and provided analysis from multiple perspectives. The experimental results show that incorporating SSL representation can improve the performance of existing speech restoration systems on various quality measures in most cases. The waveform generation networks with SSL representation exhibit robust performance for various duration of speech attenuation and better intelligibility recovery for out-of-domain data at low SNRs. Dramatic improvements in speech quality can also be found when finer local structures are provided. In the future, we plan to further investigate ways to maximize generalization capability in real recording environments.



\vfill\pagebreak
\bibliographystyle{IEEEbib}
\bibliography{strings,refs}


\end{document}